\documentclass[twoside,twocolumn,9pt]{article}
\usepackage{extsizes}
\usepackage[super,sort&compress,comma]{natbib} 
\usepackage[version=3]{mhchem}
\usepackage[left=1.5cm, right=1.5cm, top=1.785cm, bottom=2.0cm]{geometry}
\usepackage{balance}
\usepackage{mathptmx}
\usepackage{sectsty}
\usepackage{graphicx} 
\usepackage{lastpage}
\usepackage[format=plain,justification=justified,singlelinecheck=false,font={stretch=1.125,small,sf},labelfont=bf,labelsep=space]{caption}
\usepackage{float}
\usepackage{fancyhdr}
\usepackage{fnpos}
\usepackage[english]{babel}
\addto{\captionsenglish}{%
  
}
\usepackage{array}
\usepackage{droidsans}
\usepackage{charter}
\usepackage[T1]{fontenc}
\usepackage[usenames,dvipsnames]{xcolor}
\usepackage{setspace}
\usepackage[compact]{titlesec}
\usepackage{hyperref}

\definecolor{cream}{RGB}{222,217,201}

\definecolor{bleu}{RGB}{0,0,0}

\newcommand{\revisions}[1]{\textcolor{bleu}{#1}}

\RequirePackage{fix-cm}

\begin{document}

\pagestyle{fancy}
\thispagestyle{plain}
\fancypagestyle{plain}{
\renewcommand{\headrulewidth}{0pt}
}

\makeFNbottom
\makeatletter
\renewcommand\LARGE{\@setfontsize\LARGE{15pt}{17}}
\renewcommand\Large{\@setfontsize\Large{12pt}{14}}
\renewcommand\large{\@setfontsize\large{10pt}{12}}
\renewcommand\footnotesize{\@setfontsize\footnotesize{7pt}{10}}
\renewcommand\scriptsize{\@setfontsize\scriptsize{7pt}{7}}
\makeatother

\renewcommand{\thefootnote}{\fnsymbol{footnote}}
\renewcommand\footnoterule{\vspace*{1pt}%
\color{cream}\hrule width 3.5in height 0.4pt \color{black} \vspace*{5pt}} 
\setcounter{secnumdepth}{5}

\makeatletter 
\renewcommand\@biblabel[1]{#1}            
\renewcommand\@makefntext[1]%
{\noindent\makebox[0pt][r]{\@thefnmark\,}#1}
\makeatother 
\renewcommand{\figurename}{\small{Fig.}~}
\sectionfont{\sffamily\Large}
\subsectionfont{\normalsize}
\subsubsectionfont{\bf}
\setstretch{1.125} 
\setlength{\skip\footins}{0.8cm}
\setlength{\footnotesep}{0.25cm}
\setlength{\jot}{10pt}
\titlespacing*{\section}{0pt}{4pt}{4pt}
\titlespacing*{\subsection}{0pt}{15pt}{1pt}

\fancyfoot{}
\fancyfoot[LO,RE]{\vspace{-7.1pt}\includegraphics[height=9pt]{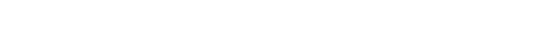}}
\fancyfoot[CO]{\vspace{-7.1pt}\hspace{13.2cm}\includegraphics{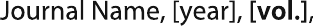}}
\fancyfoot[CE]{\vspace{-7.2pt}\hspace{-14.2cm}\includegraphics{RF}}
\fancyfoot[RO]{\footnotesize{\sffamily{1--\pageref{LastPage} ~\textbar  \hspace{2pt}\thepage}}}
\fancyfoot[LE]{\footnotesize{\sffamily{\thepage~\textbar\hspace{3.45cm} 1--\pageref{LastPage}}}}
\fancyhead{}
\renewcommand{\headrulewidth}{0pt} 
\renewcommand{\footrulewidth}{0pt}
\setlength{\arrayrulewidth}{1pt}
\setlength{\columnsep}{6.5mm}
\setlength\bibsep{1pt}

\makeatletter 
\newlength{\figrulesep} 
\setlength{\figrulesep}{0.5\textfloatsep} 

\newcommand{\topfigrule}{\vspace*{-1pt}%
\noindent{\color{cream}\rule[-\figrulesep]{\columnwidth}{1.5pt}} }

\newcommand{\botfigrule}{\vspace*{-2pt}%
\noindent{\color{cream}\rule[\figrulesep]{\columnwidth}{1.5pt}} }

\newcommand{\dblfigrule}{\vspace*{-1pt}%
\noindent{\color{cream}\rule[-\figrulesep]{\textwidth}{1.5pt}} }

\makeatother

\twocolumn[
  \begin{@twocolumnfalse}
{\includegraphics[height=30pt]{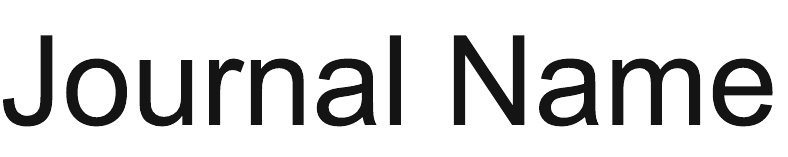}\hfill\raisebox{0pt}[0pt][0pt]{\includegraphics[height=55pt]{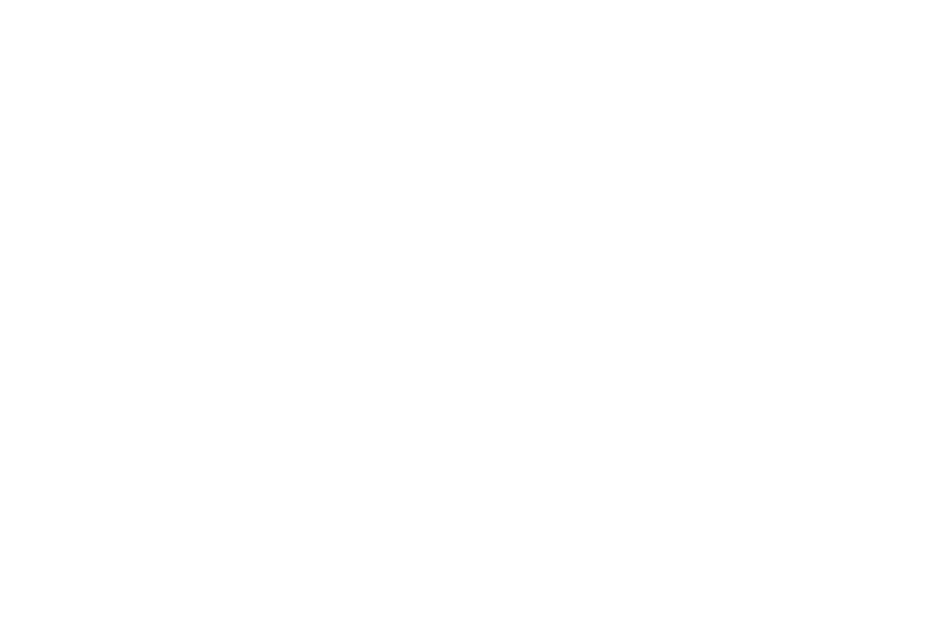}}\\[1ex]
\includegraphics[width=18.5cm]{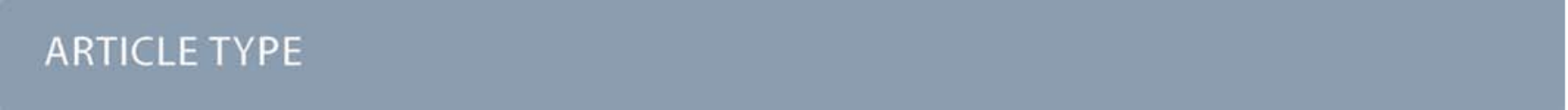}}\par
\vspace{1em}
\sffamily
\begin{tabular}{m{4.5cm} p{13.5cm} }

\includegraphics{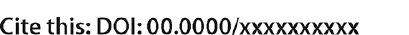} & \revisions{\noindent\LARGE{\textbf{Repulsive torques alone trigger crystallization of constant speed active particles}}} \\
 & \vspace{0.3cm} \\

 & \noindent\large{Marine Le Blay,$^{\ast}$\textit{$^{a}$} and Alexandre Morin$^{\ast}$\textit{$^{a}$}} \\

\includegraphics{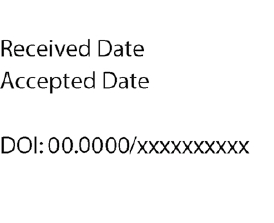} & \\

\end{tabular}

 \end{@twocolumnfalse} \vspace{0.6cm}

  ]

\renewcommand*\rmdefault{bch}\normalfont\upshape
\rmfamily
\section*{}
\vspace{-1cm}


\footnotetext{\textit{$^{a}$~Soft Matter Physics, Huygens-Kamerlingh Onnes Laboratory, Leiden University,
P.O. Box 9504, 2300 RA Leiden, The Netherlands.  E-mail: leblay@physics.leidenuniv.nl, morin@physics.leidenuniv.nl.}}

\footnotetext{\dag~Electronic Supplementary Information (ESI) available: Videos. See DOI: 00.0000/00000000.}





\sffamily{\textbf{We investigate the possibility for self-propelled particles to crystallize without reducing their intrinsic speed.
\revisions{
We illuminate how, in the absence of any force,  the competition between self-propulsion and repulsive torques determine the macroscopic phases of constant-speed active particles.
}
This minimal model expands upon existing approaches for an improved understanding of crystallization of active matter.}}\\


\rmfamily 


Can crystals form out of constituents always in motion?
Thermal systems, such as atomic and molecular matter, typically crystallize when temperature is decreased, slowing down their constituents.
This correlation between slowing down and crystallization also manifests itself in active materials~\cite{menzel2013traveling,weber2014defect,briand2016crystallization,moran2022particle}, despite their constituents being self-propelled by intrinsic forces. 
This connection even extends to other solidification processes in such materials.
A prominent example is collection of active Brownian particles which exhibit partial solidification known as motility-induced phase separation~\cite{cates2015motility,redner2013structure}.
As its name conveys, solidification in this case is entangled with speed reduction~\cite{fily2012athermal}.
Similarly, self-propelled Voronoi models~\cite{pasupalak2020hexatic,henkes2020dense} describing biological tissues exhibit jamming, another solidification transition with reduction of the cells' mobility~\cite{bi2016motility}.
All those systems share the common feature of constituents interacting through {\em forces}, inducing the reduction of their speed.
Yet,  active particles are also found to display rich collective behaviours when only interacting through {\em torques}.  
Then, only the orientation of self-propulsion is altered and their speed is preserved~\cite{chate2008collective,chepizhko2013diffusion,morin2017diffusion,bain2017critical,zhao2021phases}.
From the seminal Vicsek's model~\cite{vicsek1995novel}, those systems have proved fruitful from a theoretical perspective~\cite{toner1998flocks,peshkov2014boltzmann,solon2015pattern,kursten2020dry} and successful to account for living~\cite{bialek2012statistical} and synthetic~\cite{bricard2013emergence,morin2018flowing} systems.
Since such active systems cannot, by design,  comply with the above paradigm of solidification,  their possibility to crystallize is intriguing.
In this Communication, we prove that a minimal system of active particles interacting only {\it via} repulsive torques does crystallize.
We rationalize its fluid and crystal phases through a detailed account of binary scattering collisions.
We expect this system to complement force-based models for an ultimate comprehensive description of crystallization in out-of-equilibrium systems~\cite{bialke2012crystallization,klamser2018thermodynamic,briand2018spontaneously,digregorio20192d,bililign2021motile,reichhardt2021crystals,digregorio2022unified}, as well as to account for experimental observations~\cite{zhang2021active}.

\begin{figure}[h]
\centering
  \includegraphics[width=0.9\columnwidth]{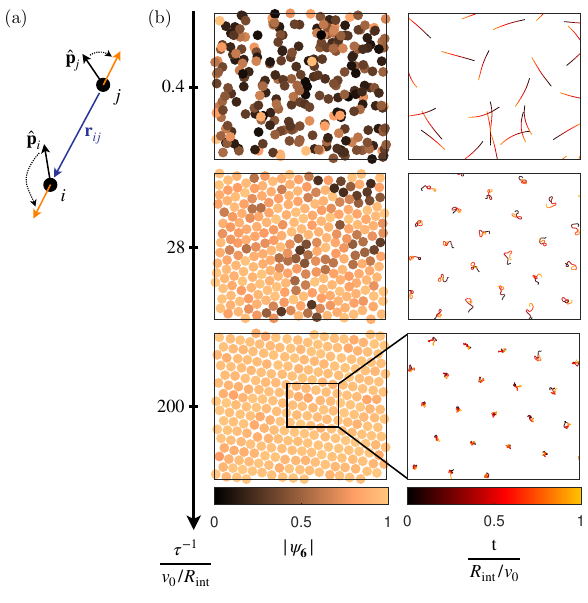}
  \caption{{\bf Liquid-to-crystal transition follows from fast re-orientation/slow self-propulsion.} 
  \revisions{(a) Schematics of the repulsive interaction between two particles.  Particles (black) re-orient so as to align (orange) opposite to their center-to-center vector (blue).
  (b)} Instantaneous configurations (left) and particles trajectories (right) increasing the repulsion over self-propulsion (top to bottom). Particles are coloured according to $|\mathbf{\psi_6}|$ evidencing the building-up of orientational order. Trajectories are all of the same duration illustrating the increasing localization of the dynamics. Dot diameter: $R_{\rm int}$. Black rectangle indicates trajectories magnification.}
  \label{Figure_Snaps}
\end{figure}
\subsection*{\bf Constant-speed, torque-only,  model system}

We introduce a minimal model of active particles moving at constant speed $v_0$ and only interacting through repulsive torques.
Particles, with instantaneous position $\mathbf{r}(t)$ and orientation $\hat{\bf{p}} (t)~= ~ \left ( \cos \theta, \sin \theta\right)$, evolve in a two-dimensional box with periodic boundaries:
\begin{equation}
\dot{\bf{r}}_i (t) = v_0 \; \hat{\bf{p}}_i (t),  \label{Eq:Motion}
\end{equation}

\begin{equation}
\dot{\theta}_i (t) = - \sum_{\left | \bf{r}_{ij}\right| < R_{\rm int}} \frac{1}{\tau} \sin \left(\theta_i - \psi_{ij} \right) \; +\; \sqrt{2 \, D_{\theta}} \, \xi_i (t), \label{Eq:Motion2}
\end{equation}

where $\mathbf{r}_{ij} = \mathbf{r}_i-\mathbf{r}_j = \left| \mathbf{r}_i-\mathbf{r}_j \right| \exp(i\psi_{ij})$, and the noise with amplitude $D_\theta$ obeys $\left \langle \xi_i (t) \right \rangle = 0$ and $\left \langle \xi_i(t) \xi_j(t') \right \rangle = \delta_{ij} \delta (t-t')$. 
We use the forward Euler method to carry out the numerical integration of Eqs.~\ref{Eq:Motion} and \ref{Eq:Motion2} with $N=270$ particles. 
Particle $i$ reorients to align in a time $\tau$ \revisions{along $\mathbf{r}_{ij}$} away from its neighbours $j$ located within an interaction range $R_{\rm int}$\revisions{, see Fig.~\ref{Figure_Snaps}a.}

Here, we focus on the competition between self-propulsion at $v_0$ and repulsion of strength $\tau^{-1}$ in the limit of small noise $D_\theta~=~0.4\times 10^{-4}\,v_0/R_{\rm int}$, thereby keeping the {\em intrinsic} persistence length constant to a value $2.5\times10^4\,R_{\rm int}$.
This competition is parametrized by the dimensionless parameter $\tau^{-1}/(v_0/R_{\rm int})$ which compares the time for a particle to travel a distance of one interaction range $R_{\rm int}/v_0$ with the time for completion of re-orientation $\tau$.
\revisions{
In what follows, we express all quantities in units of $R_{\rm int}$ for lengths and $R_{\rm int}/v_0$ for times, and study the system varying its two control parameters $\tau^{-1}$ and the number density $\rho$.}
%

\revisions{
Figure~\ref{Figure_Snaps}b and Supplementary {Videos\dag} illustrate that varying $\tau^{-1}$ at constant density $\rho ~=~1.03$ alters the nature of the macroscopic state of the system: it spontaneously transitions from a liquid state to a hexagonal crystal state.}
Indeed as $\tau^{-1}$ increases,  the system structures itself and local orientational order builds up, as shown in Fig.~\ref{Figure_Snaps}b. 
The local bond-orientational order parameter defined as $\psi_{6}^{i} \, = \left\langle e^{i 6 \psi_{ij}} \right\rangle_{j\in\mathcal{N}_i}$ goes from a broad distribution peaked around $\left | \psi_{6}  \right| \simeq 0.4$ at $\tau^{-1} = 0.4$ to a narrowly peaked distribution at $\left | \psi_{6}  \right| \simeq 1$ at $\tau^{-1}=200$\footnote[4]{The average is performed over the neighbourhood $\mathcal{N}_i$ given by Delaunay triangulation.}.
Together with this structuring, the dynamics of the constant-speed particles localize, as illustrated by the equal-duration trajectories shown in Fig.~\ref{Figure_Snaps}b.

\revisions{
In this Communication, we explore first the properties of these liquid and crystal states as $\tau^{-1}$ increases,  and then the mechanisms that shape the full two-dimensional phase diagram shown in Fig.~\ref{Figure_DiagPhases}.
}

\begin{figure}[h]
\centering
  \includegraphics[width=0.99\columnwidth]{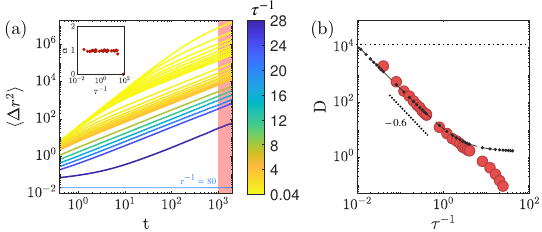}
  \caption{{\bf Dynamics in the fluid phase. } (a) Mean-squared displacement, relative to the center of mass, as a function of time for various $\tau^{-1}$ \revisions{and $\rho=1.03$} showing the diffusive dynamics characterized by the exponent $\alpha$ measured at long times (pink-shaded region), see inset. 
  The mean-squared displacements are measured at steady state from systems initialized in random configurations. 
  Thin blue line: crystal phase at $\tau^{-1}=80$.
(b) The diffusion coefficient decreases with $\tau^{-1}$ (red dots measured at long time) quantifying the slowing down of the system dynamics. Black dots: diffusion coefficient assuming only uncorrelated binary scattering collisions, showing good agreement up to $\tau^{-1} \simeq 5$.  \revisions{Dotted line: $D~=~D_{\rm noise}$.} Dashed line indicates algebraic decay with $-0.6$ slope.}
  \label{Figure_Liquid}
\end{figure}

\subsection*{\bf Fluid phase and binary collisions}
We first characterize the properties of the fluid phase occuring at low $\tau^{-1}$ \revisions{and constant $\rho = 1.03$ and} elucidate their microscopic origin.
We quantify the dynamics of the particles by measuring their mean-squared displacement $\langle \Delta r^2 \rangle(t) = \langle \left(\mathbf{r}_i(t+t_0)-\mathbf{r}_i(t_0) \right)^2 \rangle_{i,t_0}$,  shown in Fig.~\ref{Figure_Liquid}a.
For all $\tau^{-1}<28$, the dynamics is diffusive at long times: $\langle \Delta r^2\rangle~=~4D\,t^{\alpha}$ with $\alpha = 1$, see Fig.~\ref{Figure_Liquid}a inset.
The dynamics, however, slow down as $\tau^{-1}$ increases, which is captured by the decrease of the diffusion coefficient $D$ on Fig.~\ref{Figure_Liquid}b.
Note that any value of $D$ is much lower than the translational diffusion caused by the orientationnal noise of Eq.~\ref{Eq:Motion2}.
This intrinsic diffusivity $D_{\rm noise} \equiv v_0^2/(2D_\theta) = 1.25\,\times 10^4$ does not explain the liquid properties which can follow only from the interactions between particles.
Interestingly, the diffusion coefficient decays algebraically as $D\sim(\tau^{-1})^{-0.6}$, see Fig.~\ref{Figure_Liquid}b, thereby suggesting a unique origin for this decrease.

\begin{figure}[h]
\centering
  \includegraphics[width=0.9\columnwidth]{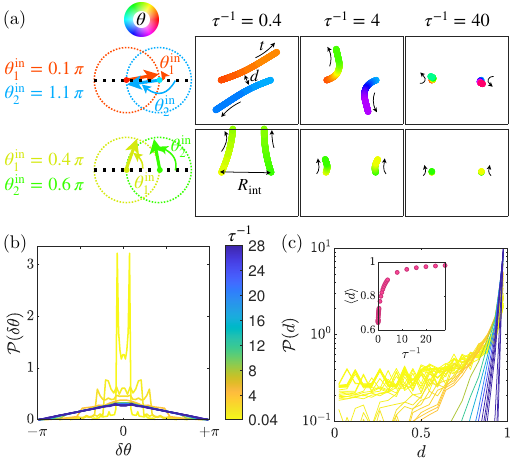}
  \caption{{\bf Binary scattering features.} (a) Trajectories of pairs of particles upon binary collisions for two sets of initial orientations $\{\theta_1^{\rm in},\,\theta_2^{\rm in}\}$ (depicted on the schematics) and three values of $\tau^{-1}$. (b) The probability density function of the scattering angles $\mathcal{P}(\delta\theta)$ broadens with $\tau^{-1}$. (c) The probability density function of the penetration depth $\mathcal{P}(d)$ reflects the strengthening of excluded-volume upon binary collisions. The inset shows the decrease of the mean value of the distribution with $\tau^{-1}$.}
  \label{Figure_Binary}
\end{figure}

A naive but simple hypothesis is that particles experience uncorrelated binary scattering events alone.
In this case, the diffusion coefficient would be given by the variance $\delta\theta^2$ of the scattering angle distribution as $D = v_0^2/(2v_0R_{\rm int}\rho\delta\theta^2)$~\cite{chepizhko2013diffusion}.
To assess the validity of this description, we investigate scattering events numerically. 
We integrate the trajectories of two particles upon collisions, with initial distance $R_{\rm int}$ and orientations $\theta_1^{\rm in}$ and $\theta_2^{\rm in}$ both uniformly distributed in $[-\pi,\,+\pi]$.
Figure~\ref{Figure_Binary}a illustrates such collisions for three different values of $\tau^{-1}$ and two sets of initial orientations.
Figure~\ref{Figure_Binary}b shows that increasing $\tau^{-1}$ broaden the distribution of scattering angles $\delta\theta = \theta_i^{\rm out} - \theta_i^{\rm in}$, with $\theta_i^{\rm out}$ the orientation after collision.
The diffusion coefficients computed from the variance of these distributions collapse on the fluid coefficients up to $\tau^{-1} \simeq 5$, as shown in Fig.~\ref{Figure_Liquid}b.
This agreement proves that the properties of the fluids are quantitatively captured by  uncorrelated binary collisions, a surprising result given their relatively high density $\rho = 1.03$.

\begin{figure}[h!]
\centering
  \includegraphics[width=0.6\columnwidth]{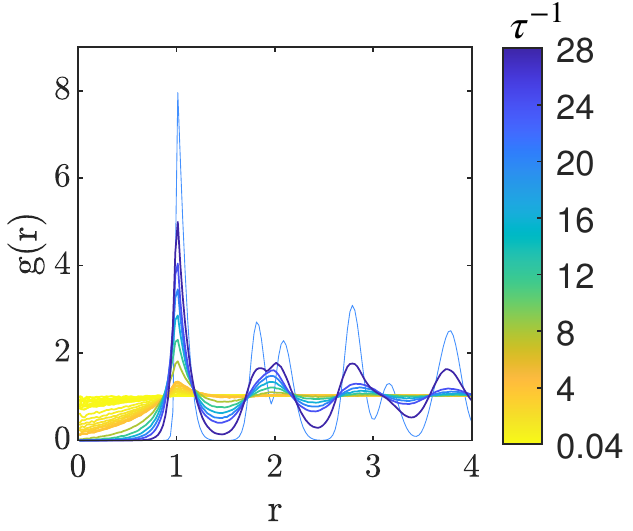}
  \caption{{\bf Structuring of the fluids.} Radial pair-correlation functions signal an evolution from a gas to a liquid state of the fluid upon increasing $\tau^{-1}$ coinciding with the enhancement of exclusion effect during binary collisions. $g(r)$ is the orthoradial average of $ g(\mathbf{r}) = \rho^{-1}\langle \sum_i \delta(\mathbf{r}-\mathbf{r}_{i}) \rangle_t $. Thin blue line: crystal phase at $\tau^{-1} = 80$.  \revisions{$\rho = 1.03$.}}
  \label{Figure_gofr}
\end{figure}

Above $\tau^{-1}>5$, however, the diffusion coefficient deviates from this simplified dynamics.
The observation of binary scattering trajectories of Fig.~\ref{Figure_Binary}a gives a hint into the mechanism at play:
as $\tau^{-1}$ increases, not only scattering is enhanced but particles also stay further apart upon colliding. 
This observation is quantified in Fig.~\ref{Figure_Binary}c through the distributions of the \revisions{minimal distance during collision $d =  \langle \min\limits_{t}(\mathbf{r}_{12}(t)) \rangle_{\{\theta^{in}_1,\theta^{in}_2\}}$.
These distributions reveal that, around $\tau^{-1} \simeq 5$,  an excluded area around the particle position emerges: they virtually never come at contact, $\mathcal{P}(d=0)~\simeq~0$.}
Importantly, this change in binary collision behaviour survives at the scale of the many-body system where depleted area forms and strengthen around each particles, as captured by the pair-correlation function $g(r)$ in Fig.~\ref{Figure_gofr}.
Driven by this exclusion effect,  the fluid evolves from an uncorrelated gas to a more and more structured liquid as $\tau^{-1}$ increases.

Altogether, the slowing down of the dynamics and the structuring of the system result from the specifics of  collisions between pairs of constant-speed particles repelling {\it via} torques.
They put the system on the path of crystallization.

\subsection*{\bf Crystal phase and effective mean-field cages}

Crystallization indeed occurs at high $\tau^{-1}$, as captured by both structural and dynamical features.
Bond-orientational order (Fig.~\ref{Figure_Snaps}) and positional order (Fig.~\ref{Figure_gofr}) extend over the whole system size.
The dynamics of the particles localize as shown by the saturation of the mean-squared displacement at $\tau^{-1}=80$ in Fig.~\ref{Figure_Liquid}a.

\begin{figure}[h]
\centering
  \includegraphics[width=0.88\columnwidth]{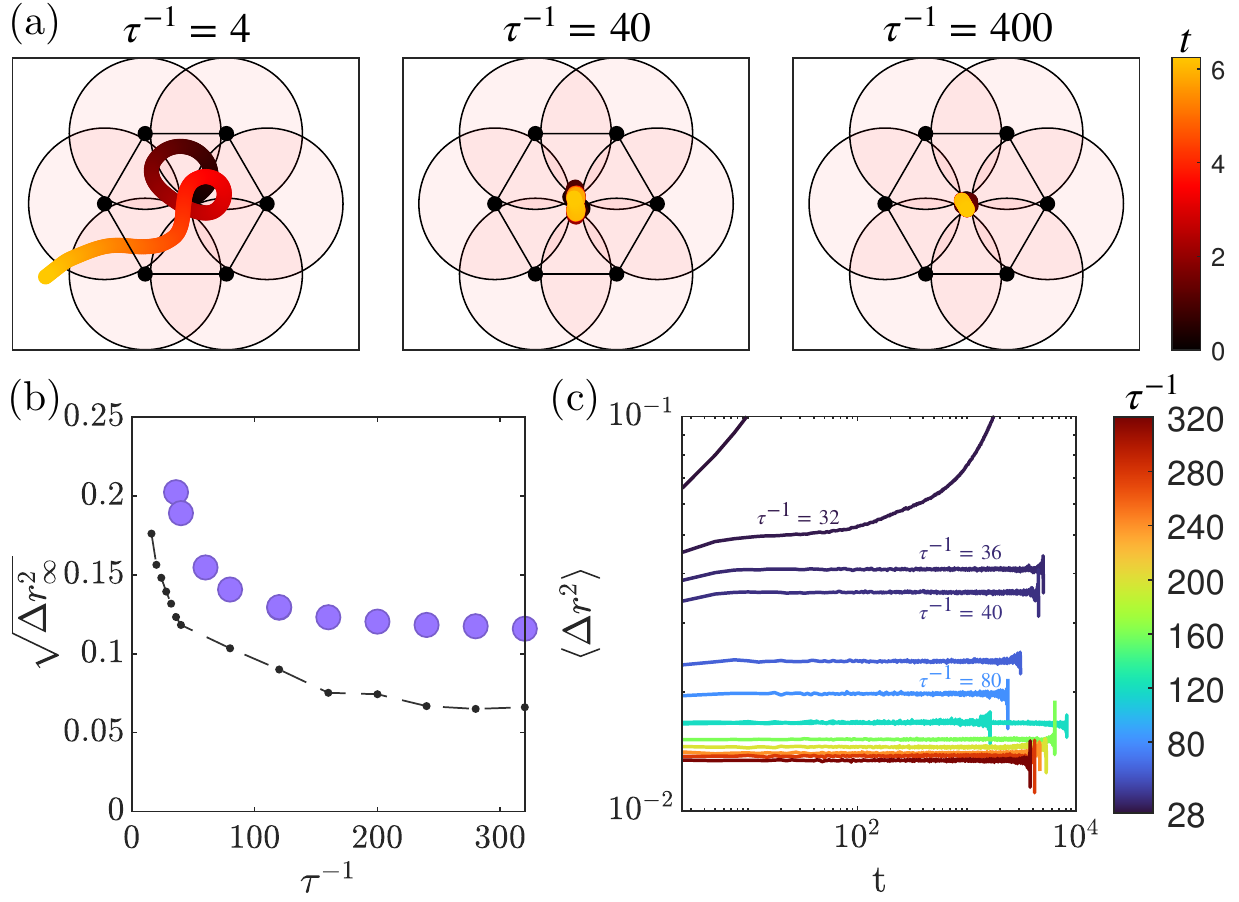}
  \caption{{\bf Efficient caging in the crystals.} (a) Six quenched particles (black dots) on a regular hexagon constitute mean-field cages. The lattice spacing respects the crystal density.  Pink circles are of radius $R_{\rm int}$ and denote the range at which the active particles, initially at the center, interact with the quenched ones.The typical trajectories (red to yellow) of constant-speed particles show two behaviours: at low$\tau^{-1}$ particles escape whereas at high $\tau^{-1}$ they are trapped.
  (b) The cage size $\sqrt{\Delta r^2_{\infty}}$ decreases with $\tau^{-1}$.  Black dots are the sizes of mean-field cages.  Purple dots are effective cage sizes in the crystals. 
(c) Mean-squared displacements, relative to the center of mass, saturate at long-time translating the localization of particles' dynamics.  The mean-squared displacements are measured at steady state from systems \revisions{at $\rho = 1.03$} initialized in crystalline configurations.  
  }
  \label{Figure_Caging}
\end{figure}

The spontaneous formation of a stable yet dynamical crystal is quite unexpected.  
Indeed,  a constant-speed particle interacting {\it via} repulsive torques with random lattices of quenched obstacles does never localize~\cite{chepizhko2013diffusion,morin2017diffusion}.
We note two main differences between the many-body system described by Eqs.~\ref{Eq:Motion} and~\ref{Eq:Motion2} and these isolated dynamics: 
Particles are interacting with (i) constantly {\em moving} particles, which, 
(ii) assemble into a {\em crystalline} lattice instead of being randomly positioned.
To elucidate which of this two differences accounts for the successful localization of the particles dynamics, we now introduce a mean-field description of the system in its crystalline state.
We investigate the dynamics of a constant-speed particle initially positioned within a mean-field unit cell consisting of a regular hexagon of quenched particles.
This configuration is depicted in Fig.~\ref{Figure_Caging}a where typical trajectories are displayed for various values of $\tau^{-1}$.
At low $\tau^{-1}$ particles escape the cell whereas at high $\tau^{-1}$ the cell act as an efficient trap.
To assess the validity of this mean-field description, we quantify the localization of the dynamics through the typical cage sizes $\Delta r^2_{\infty} = \lim\limits_{t \rightarrow\infty} \langle\Delta r^2\rangle$,  measured both from $10^3$ mean-field trajectories and from many-body systems initialized into crystalline configurations,see Fig.~\ref{Figure_Caging}b.
Overall, the mean-field description predicts well the variations of the cage sizes in the crystal; with fluctuations in the positions of particles further accounting for the slightly larger sizes in the latter case.
Besides, the threshold for efficient caging $\tau^{-1}_{\rm c} = 16$ agrees with the value yielding the loosest, possibly metastable, crystals, $\tau^{-1} = 36$, see Figs.~\ref{Figure_Caging}b-c.

Together, these results indicate that trapping by quenched crystalline lattices suffices to explain the localization of particles dynamics in the crystal phase.
In particular, no emergent correlation between particles' velocities is required for crystallization.

\subsection*{\bf Mechanisms of crystallization}
The above analyses put two mechanisms forward for the spontaneous crystallization: the increase of excluded-volume and the efficient caging within hexagonal cells.
Which of this two mechanisms sets the onset of crystallization?

\begin{figure}[h]
\centering
  \includegraphics[width=0.75\columnwidth]{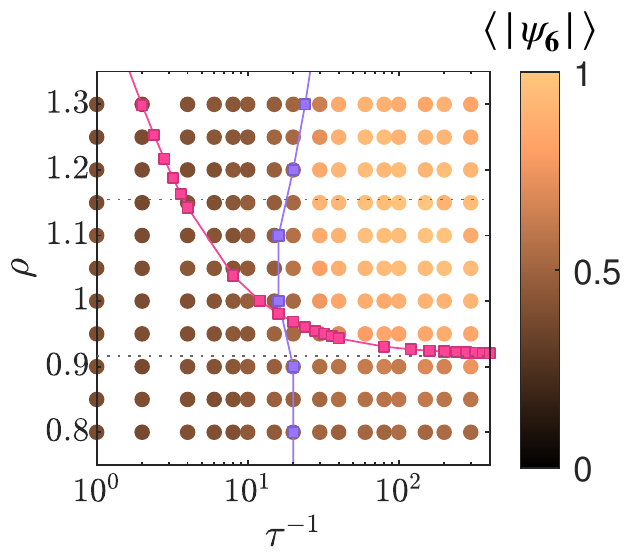}
  \caption{{\bf Necessary conditions for stable crystals. } Phase diagram of the system in the ($\rho,\,\tau^{-1}$)-plane. The dots reflect the state of the system according to the bond-orientational order parameter $\langle |\psi_6| \rangle$, measured at steady state from systems initialized in random configurations.  The crystal is confined by both criteria on excluded-volume (pink line) and on efficient caging (purple line). Dashed lines at $\rho=0.92$ and $\rho=1.15$ indicate respectively the crystallization threshold and the close-packed limit for hard-disks.}
  \label{Figure_DiagPhases}
\end{figure}

To elucidate the role of each mechanism, we systematically construct the two-dimensional phase diagram varying the density $\rho$ and the repulsion strength $\tau^{-1}$, see Fig.~\ref{Figure_DiagPhases}.
It shows that crystals emerge in a confined region of the phase diagram.

First, we rely on the measure of the mean penetration depth upon binary collision (see Fig.~\ref{Figure_Binary}c) to estimate excluded volume effects. 
Precisely, we map our system of soft particles onto hard disks of effective radius $R_{\rm eff}(\tau^{-1}) = \langle d \rangle(\tau^{-1})$ and define the effective density $\rho_{\rm eff}(\tau^{-1}) = \rho_{\rm HD}/R_{\rm eff}^2$, where $\rho_{\rm HD} \approx 0.92$ is the number density of crystallization in hard-disk systems~\cite{bernard2011two}.
Plotting the curve $\rho = \rho_{\rm eff}$ in Fig.~\ref{Figure_DiagPhases} defines a region where crystallization would occur were the excluded-volume effect to trigger the transition.
At high $\tau^{-1}  \geq 10^2$,  and upon increasing the density, crystallization occurs at $\rho_{\rm eff}$: this perfect agreement with the hard disks description suggests that entropy dominates in this region.

In stark contrast, this hard-disk ansatz does not hold at smaller $\tau^{-1}$. 
This discrepancy is elucidated by accounting for efficient caging instead.
Repeating the mean-field analysis for various densities we determine the threshold for caging $\tau^{-1}_{\rm c}(\rho)$, which varies little with $\rho$, see Fig.~\ref{Figure_DiagPhases}.
This criterion based on interactions with nearest neighbours accurately defines the onset of crystallization in this region.

Overall, the above analysis evidences that the spontaneous crystallization of the system requires that both mechanisms, excluded-volume and efficient caging, {\em simultaneously} operate.
These two simple lines of arguments, based on binary exclusion and mean-field caging, perform remarkably well to locate the crystallization region in the $(\rho,\tau^{-1})$-plane.

Considering how the crystal accommodates densities higher than closed-packed hard disks value shows how non-trivial this agreement is.
At those densities,  the crystal does not reduce its lattice spacing.
Instead some lattice positions get occupied by pairs of particles,  leading to a small reduction of $\langle |\psi_6| \rangle$,  see Fig.~\ref{Figure_DiagPhases}.
This genuine manifestation of the softness of the interactions between particles stresses the emergent character of the system crystallization.

\subsection*{\bf Conclusion}
Active particles moving at constant speed and interacting only through repulsive torques transition from fluids to crystals at high density and high repulsion strength (or, equivalently, low speeds).
The properties of the fluid phase follow from the specifics of the binary collisions: scattering sets the diffusion up until the structuring of the fluid originating from exclusion during collisions. 
While the individual particles always keep their constant speed, they crystallize through the interplay of two mechanisms: sufficient excluded volume and efficient caging by the lattice neighbours.

By demonstrating the crystallization of a minimal system of constant-speed active particles and gaining insights into the microscopic origins of its various states, we provide an approach complementary to repulsive-force systems, such as active Brownian particles, for investigating out-of-equilibrium melting. 
Refining the transition region and looking for possible hexatic~\cite{bernard2011two} or other~\cite{bialke2012crystallization} intermediate phases are sensible next steps in this direction but require accessing larger system sizes and thus appropriate computational resources.
Besides this fundamental perspective,  this simple model can serve as a starting point to build and decipher eventual experimental realizations.

\section*{Author Contributions}
M. L.B. and A. M. designed and performed the research, and wrote the manuscript.

\section*{Conflicts of interest}
There are no conflicts to declare.





\scriptsize{
\bibliography{rsc} 

\providecommand*{\mcitethebibliography}{\thebibliography}
\csname @ifundefined\endcsname{endmcitethebibliography}
{\let\endmcitethebibliography\endthebibliography}{}
\begin{mcitethebibliography}{32}
\providecommand*{\natexlab}[1]{#1}
\providecommand*{\mciteSetBstSublistMode}[1]{}
\providecommand*{\mciteSetBstMaxWidthForm}[2]{}
\providecommand*{\mciteBstWouldAddEndPuncttrue}
  {\def\EndOfBibitem{\unskip.}}
\providecommand*{\mciteBstWouldAddEndPunctfalse}
  {\let\EndOfBibitem\relax}
\providecommand*{\mciteSetBstMidEndSepPunct}[3]{}
\providecommand*{\mciteSetBstSublistLabelBeginEnd}[3]{}
\providecommand*{\EndOfBibitem}{}
\mciteSetBstSublistMode{f}
\mciteSetBstMaxWidthForm{subitem}
{(\emph{\alph{mcitesubitemcount}})}
\mciteSetBstSublistLabelBeginEnd{\mcitemaxwidthsubitemform\space}
{\relax}{\relax}

\bibitem[Menzel and L{\"o}wen(2013)]{menzel2013traveling}
A.~M. Menzel and H.~L{\"o}wen, \emph{Physical review letters}, 2013,
  \textbf{110}, 055702\relax
\mciteBstWouldAddEndPuncttrue
\mciteSetBstMidEndSepPunct{\mcitedefaultmidpunct}
{\mcitedefaultendpunct}{\mcitedefaultseppunct}\relax
\EndOfBibitem
\bibitem[Weber \emph{et~al.}(2014)Weber, Bock, and Frey]{weber2014defect}
C.~A. Weber, C.~Bock and E.~Frey, \emph{Physical review letters}, 2014,
  \textbf{112}, 168301\relax
\mciteBstWouldAddEndPuncttrue
\mciteSetBstMidEndSepPunct{\mcitedefaultmidpunct}
{\mcitedefaultendpunct}{\mcitedefaultseppunct}\relax
\EndOfBibitem
\bibitem[Briand and Dauchot(2016)]{briand2016crystallization}
G.~Briand and O.~Dauchot, \emph{Physical review letters}, 2016, \textbf{117},
  098004\relax
\mciteBstWouldAddEndPuncttrue
\mciteSetBstMidEndSepPunct{\mcitedefaultmidpunct}
{\mcitedefaultendpunct}{\mcitedefaultseppunct}\relax
\EndOfBibitem
\bibitem[Moran \emph{et~al.}(2022)Moran, Bruss, Shoenhofer, and
  Glotzer]{moran2022particle}
S.~E. Moran, I.~R. Bruss, P.~Shoenhofer and S.~C. Glotzer, \emph{Soft Matter},
  2022\relax
\mciteBstWouldAddEndPuncttrue
\mciteSetBstMidEndSepPunct{\mcitedefaultmidpunct}
{\mcitedefaultendpunct}{\mcitedefaultseppunct}\relax
\EndOfBibitem
\bibitem[Cates and Tailleur(2015)]{cates2015motility}
M.~E. Cates and J.~Tailleur, \emph{Annu. Rev. Condens. Matter Phys.}, 2015,
  \textbf{6}, 219--244\relax
\mciteBstWouldAddEndPuncttrue
\mciteSetBstMidEndSepPunct{\mcitedefaultmidpunct}
{\mcitedefaultendpunct}{\mcitedefaultseppunct}\relax
\EndOfBibitem
\bibitem[Redner \emph{et~al.}(2013)Redner, Hagan, and
  Baskaran]{redner2013structure}
G.~S. Redner, M.~F. Hagan and A.~Baskaran, \emph{Physical review letters},
  2013, \textbf{110}, 055701\relax
\mciteBstWouldAddEndPuncttrue
\mciteSetBstMidEndSepPunct{\mcitedefaultmidpunct}
{\mcitedefaultendpunct}{\mcitedefaultseppunct}\relax
\EndOfBibitem
\bibitem[Fily and Marchetti(2012)]{fily2012athermal}
Y.~Fily and M.~C. Marchetti, \emph{Physical review letters}, 2012,
  \textbf{108}, 235702\relax
\mciteBstWouldAddEndPuncttrue
\mciteSetBstMidEndSepPunct{\mcitedefaultmidpunct}
{\mcitedefaultendpunct}{\mcitedefaultseppunct}\relax
\EndOfBibitem
\bibitem[Pasupalak \emph{et~al.}(2020)Pasupalak, Yan-Wei, Ni, and
  Ciamarra]{pasupalak2020hexatic}
A.~Pasupalak, L.~Yan-Wei, R.~Ni and M.~P. Ciamarra, \emph{Soft matter}, 2020,
  \textbf{16}, 3914--3920\relax
\mciteBstWouldAddEndPuncttrue
\mciteSetBstMidEndSepPunct{\mcitedefaultmidpunct}
{\mcitedefaultendpunct}{\mcitedefaultseppunct}\relax
\EndOfBibitem
\bibitem[Henkes \emph{et~al.}(2020)Henkes, Kostanjevec, Collinson, Sknepnek,
  and Bertin]{henkes2020dense}
S.~Henkes, K.~Kostanjevec, J.~M. Collinson, R.~Sknepnek and E.~Bertin,
  \emph{Nature communications}, 2020, \textbf{11}, 1--9\relax
\mciteBstWouldAddEndPuncttrue
\mciteSetBstMidEndSepPunct{\mcitedefaultmidpunct}
{\mcitedefaultendpunct}{\mcitedefaultseppunct}\relax
\EndOfBibitem
\bibitem[Bi \emph{et~al.}(2016)Bi, Yang, Marchetti, and
  Manning]{bi2016motility}
D.~Bi, X.~Yang, M.~C. Marchetti and M.~L. Manning, \emph{Physical Review X},
  2016, \textbf{6}, 021011\relax
\mciteBstWouldAddEndPuncttrue
\mciteSetBstMidEndSepPunct{\mcitedefaultmidpunct}
{\mcitedefaultendpunct}{\mcitedefaultseppunct}\relax
\EndOfBibitem
\bibitem[Chat{\'e} \emph{et~al.}(2008)Chat{\'e}, Ginelli, Gr{\'e}goire, and
  Raynaud]{chate2008collective}
H.~Chat{\'e}, F.~Ginelli, G.~Gr{\'e}goire and F.~Raynaud, \emph{Physical Review
  E}, 2008, \textbf{77}, 046113\relax
\mciteBstWouldAddEndPuncttrue
\mciteSetBstMidEndSepPunct{\mcitedefaultmidpunct}
{\mcitedefaultendpunct}{\mcitedefaultseppunct}\relax
\EndOfBibitem
\bibitem[Chepizhko and Peruani(2013)]{chepizhko2013diffusion}
O.~Chepizhko and F.~Peruani, \emph{Physical review letters}, 2013,
  \textbf{111}, 160604\relax
\mciteBstWouldAddEndPuncttrue
\mciteSetBstMidEndSepPunct{\mcitedefaultmidpunct}
{\mcitedefaultendpunct}{\mcitedefaultseppunct}\relax
\EndOfBibitem
\bibitem[Morin \emph{et~al.}(2017)Morin, Cardozo, Chikkadi, and
  Bartolo]{morin2017diffusion}
A.~Morin, D.~L. Cardozo, V.~Chikkadi and D.~Bartolo, \emph{Physical Review E},
  2017, \textbf{96}, 042611\relax
\mciteBstWouldAddEndPuncttrue
\mciteSetBstMidEndSepPunct{\mcitedefaultmidpunct}
{\mcitedefaultendpunct}{\mcitedefaultseppunct}\relax
\EndOfBibitem
\bibitem[Bain and Bartolo(2017)]{bain2017critical}
N.~Bain and D.~Bartolo, \emph{Nature communications}, 2017, \textbf{8},
  1--6\relax
\mciteBstWouldAddEndPuncttrue
\mciteSetBstMidEndSepPunct{\mcitedefaultmidpunct}
{\mcitedefaultendpunct}{\mcitedefaultseppunct}\relax
\EndOfBibitem
\bibitem[Zhao \emph{et~al.}(2021)Zhao, Ihle, Han, Huepe, and
  Romanczuk]{zhao2021phases}
Y.~Zhao, T.~Ihle, Z.~Han, C.~Huepe and P.~Romanczuk, \emph{Physical Review E},
  2021, \textbf{104}, 044605\relax
\mciteBstWouldAddEndPuncttrue
\mciteSetBstMidEndSepPunct{\mcitedefaultmidpunct}
{\mcitedefaultendpunct}{\mcitedefaultseppunct}\relax
\EndOfBibitem
\bibitem[Vicsek \emph{et~al.}(1995)Vicsek, Czir{\'o}k, Ben-Jacob, Cohen, and
  Shochet]{vicsek1995novel}
T.~Vicsek, A.~Czir{\'o}k, E.~Ben-Jacob, I.~Cohen and O.~Shochet, \emph{Physical
  review letters}, 1995, \textbf{75}, 1226\relax
\mciteBstWouldAddEndPuncttrue
\mciteSetBstMidEndSepPunct{\mcitedefaultmidpunct}
{\mcitedefaultendpunct}{\mcitedefaultseppunct}\relax
\EndOfBibitem
\bibitem[Toner and Tu(1998)]{toner1998flocks}
J.~Toner and Y.~Tu, \emph{Physical review E}, 1998, \textbf{58}, 4828\relax
\mciteBstWouldAddEndPuncttrue
\mciteSetBstMidEndSepPunct{\mcitedefaultmidpunct}
{\mcitedefaultendpunct}{\mcitedefaultseppunct}\relax
\EndOfBibitem
\bibitem[Peshkov \emph{et~al.}(2014)Peshkov, Bertin, Ginelli, and
  Chat{\'e}]{peshkov2014boltzmann}
A.~Peshkov, E.~Bertin, F.~Ginelli and H.~Chat{\'e}, \emph{The European Physical
  Journal Special Topics}, 2014, \textbf{223}, 1315--1344\relax
\mciteBstWouldAddEndPuncttrue
\mciteSetBstMidEndSepPunct{\mcitedefaultmidpunct}
{\mcitedefaultendpunct}{\mcitedefaultseppunct}\relax
\EndOfBibitem
\bibitem[Solon \emph{et~al.}(2015)Solon, Caussin, Bartolo, Chat{\'e}, and
  Tailleur]{solon2015pattern}
A.~P. Solon, J.-B. Caussin, D.~Bartolo, H.~Chat{\'e} and J.~Tailleur,
  \emph{Physical Review E}, 2015, \textbf{92}, 062111\relax
\mciteBstWouldAddEndPuncttrue
\mciteSetBstMidEndSepPunct{\mcitedefaultmidpunct}
{\mcitedefaultendpunct}{\mcitedefaultseppunct}\relax
\EndOfBibitem
\bibitem[K{\"u}rsten and Ihle(2020)]{kursten2020dry}
R.~K{\"u}rsten and T.~Ihle, \emph{Physical Review Letters}, 2020, \textbf{125},
  188003\relax
\mciteBstWouldAddEndPuncttrue
\mciteSetBstMidEndSepPunct{\mcitedefaultmidpunct}
{\mcitedefaultendpunct}{\mcitedefaultseppunct}\relax
\EndOfBibitem
\bibitem[Bialek \emph{et~al.}(2012)Bialek, Cavagna, Giardina, Mora, Silvestri,
  Viale, and Walczak]{bialek2012statistical}
W.~Bialek, A.~Cavagna, I.~Giardina, T.~Mora, E.~Silvestri, M.~Viale and A.~M.
  Walczak, \emph{Proceedings of the National Academy of Sciences}, 2012,
  \textbf{109}, 4786--4791\relax
\mciteBstWouldAddEndPuncttrue
\mciteSetBstMidEndSepPunct{\mcitedefaultmidpunct}
{\mcitedefaultendpunct}{\mcitedefaultseppunct}\relax
\EndOfBibitem
\bibitem[Bricard \emph{et~al.}(2013)Bricard, Caussin, Desreumaux, Dauchot, and
  Bartolo]{bricard2013emergence}
A.~Bricard, J.-B. Caussin, N.~Desreumaux, O.~Dauchot and D.~Bartolo,
  \emph{Nature}, 2013, \textbf{503}, 95--98\relax
\mciteBstWouldAddEndPuncttrue
\mciteSetBstMidEndSepPunct{\mcitedefaultmidpunct}
{\mcitedefaultendpunct}{\mcitedefaultseppunct}\relax
\EndOfBibitem
\bibitem[Morin and Bartolo(2018)]{morin2018flowing}
A.~Morin and D.~Bartolo, \emph{Physical Review X}, 2018, \textbf{8},
  021037\relax
\mciteBstWouldAddEndPuncttrue
\mciteSetBstMidEndSepPunct{\mcitedefaultmidpunct}
{\mcitedefaultendpunct}{\mcitedefaultseppunct}\relax
\EndOfBibitem
\bibitem[Bialk{\'e} \emph{et~al.}(2012)Bialk{\'e}, Speck, and
  L{\"o}wen]{bialke2012crystallization}
J.~Bialk{\'e}, T.~Speck and H.~L{\"o}wen, \emph{Physical review letters}, 2012,
  \textbf{108}, 168301\relax
\mciteBstWouldAddEndPuncttrue
\mciteSetBstMidEndSepPunct{\mcitedefaultmidpunct}
{\mcitedefaultendpunct}{\mcitedefaultseppunct}\relax
\EndOfBibitem
\bibitem[Klamser \emph{et~al.}(2018)Klamser, Kapfer, and
  Krauth]{klamser2018thermodynamic}
J.~U. Klamser, S.~C. Kapfer and W.~Krauth, \emph{Nature communications}, 2018,
  \textbf{9}, 1--8\relax
\mciteBstWouldAddEndPuncttrue
\mciteSetBstMidEndSepPunct{\mcitedefaultmidpunct}
{\mcitedefaultendpunct}{\mcitedefaultseppunct}\relax
\EndOfBibitem
\bibitem[Briand \emph{et~al.}(2018)Briand, Schindler, and
  Dauchot]{briand2018spontaneously}
G.~Briand, M.~Schindler and O.~Dauchot, \emph{Physical review letters}, 2018,
  \textbf{120}, 208001\relax
\mciteBstWouldAddEndPuncttrue
\mciteSetBstMidEndSepPunct{\mcitedefaultmidpunct}
{\mcitedefaultendpunct}{\mcitedefaultseppunct}\relax
\EndOfBibitem
\bibitem[Digregorio \emph{et~al.}(2019)Digregorio, Levis, Suma, Cugliandolo,
  Gonnella, and Pagonabarraga]{digregorio20192d}
P.~Digregorio, D.~Levis, A.~Suma, L.~F. Cugliandolo, G.~Gonnella and
  I.~Pagonabarraga, Journal of Physics: Conference Series, 2019, p.
  012073\relax
\mciteBstWouldAddEndPuncttrue
\mciteSetBstMidEndSepPunct{\mcitedefaultmidpunct}
{\mcitedefaultendpunct}{\mcitedefaultseppunct}\relax
\EndOfBibitem
\bibitem[Bililign \emph{et~al.}(2021)Bililign, Balboa~Usabiaga, Ganan, Poncet,
  Soni, Magkiriadou, Shelley, Bartolo, and Irvine]{bililign2021motile}
E.~S. Bililign, F.~Balboa~Usabiaga, Y.~A. Ganan, A.~Poncet, V.~Soni,
  S.~Magkiriadou, M.~J. Shelley, D.~Bartolo and W.~Irvine, \emph{Nature
  Physics}, 2021,  1--7\relax
\mciteBstWouldAddEndPuncttrue
\mciteSetBstMidEndSepPunct{\mcitedefaultmidpunct}
{\mcitedefaultendpunct}{\mcitedefaultseppunct}\relax
\EndOfBibitem
\bibitem[Reichhardt and Reichhardt(2021)]{reichhardt2021crystals}
C.~J. Reichhardt and C.~Reichhardt, \emph{Nature Physics}, 2021,  1--2\relax
\mciteBstWouldAddEndPuncttrue
\mciteSetBstMidEndSepPunct{\mcitedefaultmidpunct}
{\mcitedefaultendpunct}{\mcitedefaultseppunct}\relax
\EndOfBibitem
\bibitem[Digregorio \emph{et~al.}(2022)Digregorio, Levis, Cugliandolo,
  Gonnella, and Pagonabarraga]{digregorio2022unified}
P.~Digregorio, D.~Levis, L.~F. Cugliandolo, G.~Gonnella and I.~Pagonabarraga,
  \emph{Soft Matter}, 2022, \textbf{18}, 566--591\relax
\mciteBstWouldAddEndPuncttrue
\mciteSetBstMidEndSepPunct{\mcitedefaultmidpunct}
{\mcitedefaultendpunct}{\mcitedefaultseppunct}\relax
\EndOfBibitem
\bibitem[Zhang \emph{et~al.}(2021)Zhang, Alert, Yan, Wingreen, and
  Granick]{zhang2021active}
J.~Zhang, R.~Alert, J.~Yan, N.~S. Wingreen and S.~Granick, \emph{Nature
  Physics}, 2021, \textbf{17}, 961--967\relax
\mciteBstWouldAddEndPuncttrue
\mciteSetBstMidEndSepPunct{\mcitedefaultmidpunct}
{\mcitedefaultendpunct}{\mcitedefaultseppunct}\relax
\EndOfBibitem
\bibitem[Bernard and Krauth(2011)]{bernard2011two}
E.~P. Bernard and W.~Krauth, \emph{Physical review letters}, 2011,
  \textbf{107}, 155704\relax
\mciteBstWouldAddEndPuncttrue
\mciteSetBstMidEndSepPunct{\mcitedefaultmidpunct}
{\mcitedefaultendpunct}{\mcitedefaultseppunct}\relax
\EndOfBibitem
\end{mcitethebibliography}
\bibliographystyle{rsc} } 

\end{document}